
\input harvmac
\def\half{{1 \over 2}}
\def\dzm{{\partial_z}}
\def\dz{{\partial_z}}

\def\dzbar{{\bar\partial _{\bar z}}}
\def\jb{{\bar j}}
\def\pj{{\partial _j}}
\def\pjb{{\bar\partial^j}}

\def\N{{\nabla}}
\def\Nb{{\bar\nabla}}
\def\pb{{\bar p}}

\def\pbj{{\bar\partial _{\bar j}}}

\def\L{{\Lambda}}
\def\Gf{{{\cal{G}}_4^+}}
\def\Gs{{{\cal{G}}_6^+}}
\def\Gtf{{\tilde{\cal{G}}_4^+}}
\def\Gts{{\tilde{\cal{G}}_6^+}}
\def\P{{\Phi}}

\def\ep {{\epsilon^{jk}}}
\def\epb {{\epsilon^{\bar j\bar k}}}
\def\xj {{x_j}}
\def\xk {{x_k}}
\def\xbj {{\bar x_{\bar j}}}
\def\xbk {{\bar x_{\bar k}}}

\def\a {{\alpha}}
\def\b {{\beta}}
\def\g {{\gamma}}

\def\ad {{\dot\alpha}}

\def\k {{\kappa}}

\def\t {{\theta}}
\def\ta {{\theta^\alpha}}

\def\tba {{\bar\theta^\ad}}
\def\oj{{\omega_j}}
\def\obj{{\bar\omega^j}}

\def\obk{{\bar\omega^k}}

\def\obl{{\bar\omega^l}}
\def\O{{\Omega}}
\def\Ob{{\bar\Omega}}

\def \ad {{\dot \a}}

\def \t {{\theta}}
\def \tb {{\bar\theta}}
\def \Gp {{G^+}}
\def \Gtp {{\tilde G^+}}

\def \Gpf {{G^+_4}}
\def \Gtpf {{\tilde G^+_4}}
\def \Gps {{G^+_6}}
\def \Gtps {{\tilde G^+_6}}
\Title{\vbox{\hbox{IFUSP-P-1143}}}
{\vbox{\centerline{\bf Super-Poincar\'e Invariant Superstring Field
Theory}}}
\bigskip\centerline{Nathan Berkovits}
\bigskip\centerline{Dept. de F\'{\i}sica Matem\'atica, Univ. de S\~ao Paulo}
\centerline{CP 20516, S\~ao Paulo, SP 01498, BRASIL}
\centerline{and}
\centerline{IMECC, Univ. de Campinas}
\centerline{CP 1170, Campinas, SP 13100, BRASIL}
\bigskip\centerline{e-mail: nberkovi@snfma1.if.usp.br}
\vskip .2in

   Using the topological techniques developed in an earlier paper with
Vafa, a field theory action is constructed for any open string
with critical N=2 worldsheet superconformal invariance. Instead of
the Chern-Simons-like action found by Witten, this action resembles
that of a Wess-Zumino-Witten model. For the N=2 string which
describes (2,2) self-dual Yang-Mills, the string field generalizes
the scalar field of Yang.

    As was shown in recent papers, an N=2 string can also be used
to describe the Green-Schwarz superstring in a Calabi-Yau background.
In this case, one needs three types of string fields which generalize
the real superfield of the super-Yang-Mills prepotential, and the chiral
and anti-chiral superfields of the Calabi-Yau scalar multiplet. The
resulting field theory action for the open superstring in a Calabi-Yau
background has the advantages over the standard RNS action that it
is manifestly SO(3,1) super-Poincar\'e invariant and does not require
contact terms to remove tree-level divergences.

\Date{March 1995}
\newsec {Introduction}

The construction of a field theory action for the superstring is
an important problem since it may lead to clues about
non-perturbative superstring theory which are unobtainable from the
on-shell perturbative S-matrix. The standard approach to constructing
such an action makes use of the N=1 BRST operator $Q$, together with
the picture-changing operator $Z$ and the inverse-picture-changing
operator $Y$, of the RNS formalism.\ref\wsup{E. Witten, Nucl. Phys. B268
(1986) 253\semi E. Witten, Nucl. Phys. B276 (1986) 291.}

The naive equation of motion and gauge-invariance for the
Neveu-Schwarz string field $A$ is
$Q A +Z A^2=0$
and $\delta A=Q \Lambda +Z [A, \Lambda]$,
where string fields are multiplied using Witten's half-string product and
$Z$ is inserted at the vertex midpoint.
This naive equation of motion can be obtained from the
Cherns-Simons-like action
$\int (A Q A +{2\over 3}
Z A^3),$
however gauge invariance requires the addition of quartic
and higher-order contact terms\ref\conw{C. Wendt, Nucl. Phys. B314 (1989) 209.}
to remove the divergence when
two $Z$'s collide.\foot{
Some authors\ref\pr{C.R. Preitschopf, C.B. Thorn, and S. Yost,
Nucl. Phys. B337 (1990) 363\semi
I. Ya. Aref'eva, P.B. Medvedev, and A.P. Zubarev,
Nucl. Phys. B341 (1990) 464.}
have tried to avoid these divergences by modifying
the picture of $A$ so that the action takes the form
$\int (A Q Y^2 A +{2\over 3}Y^2 A^3)$.
However this action suffers from the problem that the linearized
equation of motion, $Q Y^2 A=0$, has unphysical solutions
when $Y^2 A=0$.
\ref\ya{I. Ya. Aref'eva and P.B. Medvedev,
Phys. Lett. 202B (1988) 510.}}

Another disadvantage of the RNS superstring field theory is
the lack of manifest spacetime supersymmetry. Besides requiring
different string fields for the bosonic and fermionic sectors,
the action explicitly involves the picture-changing operators,
$Z$ and $Y$, which do not commute with the spacetime-supersymmetry
generators of picture $\pm\half.$

A different approach to constructing a field theory action for the
superstring uses the light-cone Green-Schwarz formalism.\ref\gs
{M.B. Green and J.H. Schwarz, Nucl. Phys. B243 (1984) 475.} This
action is manifestly invariant under an SU(4)$\times$U(1)
subgroup of the super-Poincar\'e group. However because it is
completely gauge-fixed, it is difficult to find a geometrical
structure. Furthermore, the light-cone Green-Schwarz field theory
action suffers from the same problem as the RNS action that
contact terms need to be introduced to remove tree-level
divergences.\ref\k{J. Greensite and F.R. Klinkhamer, Nucl. Phs. B281
(1987) 269.}

Over the last few years, a new formalism has been developed for the
superstring which has critical N=2 worldsheet superconformal
invariance.\ref\me{N. Berkovits, Nucl. Phys. B341 (1994) 258.}
This formalism is manifestly SO(3,1) super-Poincar\'e
invariant and can be used to describe any compactified version
of the superstring which contains four-dimensional
spacetime-supersymmetry.

Because of the problems with other formalisms, it is natural
to try to construct a superstring field theory action using this
new formalism. The most obvious approach would be to use the
N=2 worldsheet superconformal invariance to construct an N=2
BRST operator and N=2 picture-changing operators, and to look
for an N=2 generalization of the Chern-Simons-like action. However
this approach would probably suffer from the same
contact term problems as the other formalisms.

A second approach is to twist the N=2 superconformal generators
using the topological techniques developed with Vafa in reference
\ref\vafa{N. Berkovits and C. Vafa, Nucl. Phys. B433 (1995) 123.}, and to
construct a new type of string field theory action. As will
be shown in this paper, this new type of action resembles a
Wess-Zumino-Witten action \ref\wzw{E. Witten,
Comm. Math. Phys. 92 (1984) 455}, rather than a Chern-Simons action.
The two major advantages of this action are that
it is manifestly SO(3,1) super-Poincar\'e invariant and that
it does not require contact terms to remove tree-level divergences.

In section 2 of this paper, the topological description \vafa of
critical N=2 strings
will be reviewed. In this topological description, the BRST operator
$Q$ is replaced by two fermionic spin-one generators, $\Gp$ and
$\Gtp$, which are constructed entirely out of N=2 matter fields.
The linearized equation of motion
$Q A=0$ is replaced by $\Gtp\Gp\Phi=0$ and the linearized gauge invariance
$\delta A=Q\Lambda$ is replaced by $\delta \Phi=$
$\Gp\Lambda+\Gtp\bar\Lambda$.

In section 3, it will be shown how to construct a string field
theory action from this topological description of critical N=2
strings. Whereas $Q$ is covariantized to $Q+A$ in the
Chern-Simons-like action, $\Gp$ and $\Gtp$ will be covariantized
to $e^{-\Phi} \Gp e^{\Phi}$ and $\Gtp$. So instead of the
non-linear equation of motion $Q A +A^2=0$,
one finds the equation of motion
$\Gtp(e^{-\Phi}\Gp e^{\Phi})=0$, which comes from
a Wess-Zumino-Witten type of action.
For the critical N=2 string that describes (2,2) self-dual
Yang-Mills, it is easy to show that the resulting string
field theory action reproduces the action for self-dual
Yang-Mills\ref\ov{H. Ooguri and C. Vafa, Nucl. Phys. B367 (1991) 83.}
where the string field
generalizes the scalar field of
Yang\ref\Yang{C.N. Yang, Phys. Rev. Lett. 38 (1977) 1377.}.

In section 4, the N=2 description of the superstring
in a Calabi-Yau background will be reviewed. This N=2 description
is obtained by embedding the critical N=1 superstring into a
critical N=2 string such that $\Gp$ is the integrand of the N=1 BRST
operator and $\Gtp$ is the $\eta$ fermion which comes from fermionizing
the N=1 ghosts.\ref\bv{N. Berkovits, Nucl. Phys. B420 (1994) 332\semi
N. Berkovits and C. Vafa, Mod. Phys. Lett. A9 (1994) 653.}
N=2 vertex operators are related to N=1
vertex operators by $\Phi=\xi A$, so $\Gtp\Gp\Phi$=0 implies $QA=0$.
The main advantage of this N=2 description of the superstring is
it can be made manifestly super-Poincar\'e invariant by expressing
the RNS variables in terms of GS-like variables.
These GS-like
variables include the four-dimensional superspace variables
$x^m$, $\theta^\a$ and $\bar\theta^\ad$, the conjugate variables
for
$\theta^\a$ and $\bar\theta^\ad$, the chiral boson $\rho$ (similar
to the chiral boson $\phi$ of the N=1 ghost sector), and any
$c=9$ N=2 superconformal field theory.\me

In section 5, a manifestly super-Poincar\'e invariant field theory
action will be constructed for the superstring. Although this might
seem straightforward using results from the previous sections, there is
one difficulty which needs to be overcome. Since the linearized equation
of motion $\Gtp\Gp\Phi$ has solutions at every N=1 picture, the
same physical state can be represented by different string fields.
However it will be proven that if the string field is restricted to
have $\rho$-charge $+1,$ 0, or $-1$, each physical state
is uniquely represented.
The string field with zero $\rho$-charge will generalize
the real superfield that describes the super-Yang-Mills prepotential,
while the string fields with $\pm 1$ $\rho$-charge will generalize
the chiral and anti-chiral superfields that describe the Calabi-Yau
scalar multiplet. In terms of these three string fields, equations of
motion and gauge invariances will be expressed in a manifestly
super-Poincar\'e invariant manner.

Miraculously, an action which yields these equations of motion can be found
by comparing with the action of Marcus, Sagnotti, and Siegel for
ten-dimensional super-Yang-Mills written in terms of four-dimensional
superfields.\ref\mar{N. Marcus, A. Sagnotti, and W. Siegel,
Nucl. Phys. B224 (1983) 159.}
By promoting their point-particle superfields to
string fields, a manifestly SO(3,1) super-Poincar\'e invariant
action will be constructed for any compactification of the open superstring
which preserves four-dimensional spacetime-supersymmetry.
This superstring field theory action resembles a WZW action and does
not require contact terms to remove tree-level divergences.

In section 6, the conclusions of this paper will be summarized and some
possible applications will be proposed.

\newsec {Review of the Topological Description of Critical N=2 Strings}

A critical N=2 string contains generators $L$,
$G^+$, $G^-$, and $J$ which satisfy the OPE's of an N=2
superconformal algebra with $c=6$. Since
the OPE of $J(y)$ with $J(z)$ goes like $2 (y-z)^{-2}$, this N=2
algebra can be extended to a small N=4 superconformal algebra
by defining two new spin-one generators, $J^{++}=e^{\int^z J}$ and
$J^{--}=e^{-\int^z J}$,
and two new spin-3/2 generators, $\Gtp$ and
$\tilde G^-$, which are defined by taking the pole term
in the OPE of $J^{++}$ with $G^-$ and $J^{--}$ with $G^+$.
Note that $G^+$ has no singularities with $\tilde G^-$,
and $\int G^+$ anticommutes with $\int \Gtp$ since
the OPE of $G^+(y)$ with $\Gtp(z)$ goes like
$\half(\partial_z -\partial_y)((y-z)^{-1}J^{++}(z))$.

One method for calculating scattering amplitudes for the
critical N=2 string is to introduce N=2 ghosts, construct an N=2
BRST operator, and integrate correlation functions of BRST-invariant
vertex operators over the moduli of N=2 super-Riemann surfaces.
However as was shown in a recent paper with Vafa \vafa, there is a
simpler method for calculating scattering amplitudes. After
twisting the algebra by the U(1) current $J$ such that
$\Gp$ and $\Gtp$ have spin-one while $G^-$ and $\tilde G^-$ have
spin-two, one can find a convenient choice of N=2 moduli such that
the N=2 ghost correlation functions cancel each other out.

After integrating out the N=2 ghost fields, the scattering amplitude
can be expressed as an integral over ordinary N=0 moduli of
correlation functions of ghost-independent vertex operators. Instead
of being contracted with $b$ ghosts, the
$3g-3+N$ beltrami differentials for the moduli are contracted with
the spin-two $G^-$'s and $\tilde G^-$'s (the relative number of
$G^-$'s and $\tilde G^-$'s depends on the U(1) instanton number of the
original N=2 super-Riemann surface).

Although amplitudes for arbitrary genus and U(1) instanton number
can be found in reference \vafa, the most relevant amplitude
for open string field
theory is the three-point tree amplitude at zero instanton number
which is given by:
\eqn\three{<\Phi(z_1) (G^+\Phi(z_2))(\tilde G^+\Phi(z_3))>}
where $\Phi$ is a U(1)-neutral vertex operator satisfying
$\Gtp\Gp\Phi=0$ (for unitary strings, this implies that $\Phi$ is
a weight-zero N=2 primary field), and $G^+ \Phi$ signifies the
contour integral of spin-one $\Gp$ around $\Phi$. As in all open
string theories, the $\Phi$'s carry Chan-Paton factors which will
be supressed throughout this paper.

By deforming the contours of $G^+$ and $\Gtp$, it is easy to check
that this amplitude is symmetric in the three vertices and is
invariant under the gauge transformations
$\delta\Phi=G^+\Lambda +\Gtp\bar\Lambda$. Note that the contour
integral of $G^+$ anticommutes with the contour integral of $\Gtp$.

The simplest example of a critical N=2 string is the open string
which describes (2,2) self-dual Yang-Mills.\ov Its action is
\eqn\two{\int dz d\bar z (\dz x_j \dzbar \bar x_{\bar j} +
\psi^-_{\bar j} \dzbar \psi^+_j
+\bar\psi^-_{\bar j} \dz \bar\psi^+_j),}
and its N=2 superconformal generators are
$$L=\dzm x_j \dzm \xbj +\psi^-_\jb\dzm\psi^+_j,\quad
G^+ =\psi^+_j\dzm\xbj,\quad
G^- =\psi^-_\jb\dzm\xj,\quad
J=\psi^+_j\psi^-_\jb$$
where $j$ and $\bar j$ take the values 1 or 2.
Note that after twisting,
$\psi^+_j$ is spin-zero while $\psi^-_\jb$ is spin-one.
The additional N=4 superconformal generators are easily found to be
$$\tilde G^+ =\ep \psi^+_j\dzm\xk,\quad
\tilde G^- =\epb \psi^-_\jb\dzm\xbk,\quad
J^{++}=\ep\psi^+_j\psi^+_k,\quad
J^{--}=\epb\psi^-_j\psi^-_k.$$

Up to gauge transformations, the only momentum-dependent U(1)-neutral
vertex operator satisfying $\Gtp\Gp\Phi=0$ is
$\Phi=\exp(i k_j \bar x_{\bar j}+i \bar k_{\bar j} x_j)$ where
$k_j \bar k_{\bar j}=0.$ After performing the correlation functions
over the N=2 matter fields (note that $\psi^+_j$ has a zero mode),
one finds that \three produces the usual three-point tree amplitude
$\bar k_{\bar j}^2 k_j^3 f^{I_1 I_2 I_3}$ where
$f^{I_1 I_2 I_3}$ is the structure constant for the Chan-Paton
factors.

\newsec {String Field Theory for Open N=2 Strings}

In order to construct a string field theory action, it is natural
to generalize the on-shell vertex operator to an off-shell string
field $\Phi$ which is a
U(1)-neutral function of
$x(\sigma)$ and $\psi(\sigma)$ (note that $\Phi$ is bosonic,
in contrast with the Neveu-Schwarz string field, $A$, which is fermionic).
Since complex conjugation flips $J \to -J$ and therefore changes the
sign of the twist, the reality condition on $\Phi$ will be
\eqn\conj{(\overline{\Phi(\sigma)})^R=\Phi(\pi-\sigma)}
where the bar signifies
hermitian conjugation, and the $R$ signifies an SU(2) rotation which
returns the original twist by transforming $J\to -J$, $J^{++}\to J^{--}$,
and $J^{--}\to J^{++}$. For an N=2 primary field $\Psi$
of U(1)-charge $m$,
$(\Psi)^R$ is the pole of order $m^2$ in the OPE of $\Psi$ with
$e^{-m\int^z J}$ (e.g., for the self-dual N=2 string,
$(\psi_j^+)^R=\epsilon_{jk}\psi_{\bar k}^-$).

For the string field theory action to be correct,
the quadratic term in the
action should enforce the linearized
equation of motion $\Gtp\Gp\Phi=0$, while the cubic term should
produce the correct on-shell three-point amplitude. Finally, the action
should contain a gauge invariance whose linearized form is
$\delta\Phi=\Gp\Lambda +\Gtp \bar\Lambda$.

The quadratic and cubic terms in the action are easily found to be
of the form
$$\int\left( (\Gp\Phi)(\Gtp\Phi)+\Phi\{\Gp\Phi,\Gtp\Phi\}\right) ,$$
however there is no non-linear version of the gauge transformation
$\delta\Phi=\Gp\Lambda+\Gtp\bar\Lambda$ which leaves this action
invariant. This means that quartic and higher-order terms need to be added,
which should not be surprising since the non-linear equation of motion for
(2,2) self-dual Yang-Mills is $\pj(e^{-\phi}\pbj
e^{\phi})=0$, where
$A_j= e^{\half\phi}\pj e^{-\half\phi}$ and
$A_{\bar j}= e^{-\half\phi}\pbj e^{\half\phi}$ are the
self-dual Yang-Mills gauge fields\Yang. Note however that the quartic and
higher-order terms will not have infinite coefficients like those of
the RNS field theory action, and will be completely explicit functions
of the string fields.

The obvious guess for the non-linear generalization of $\Gtp\Gp\Phi=0$
is therefore $\Gtp(e^{-\Phi}\Gp e^{\Phi})=0$ where multiplication
of string fields is always performed using Witten's half-string
overlap. If $\phi$ is the component
of $\Phi$ which depends only on the zero mode of $x$, then
$\Gtp(e^{-\Phi}\Gp e^{\Phi})$ contains the term
$\epsilon^{kl}\psi_k^+ \partial_l
(e^{-\phi}\psi^+_j\pbj e^{\phi})$=
$(\half\epsilon^{kl}\psi^+_k\psi^+_l)
\partial_j(e^{-\phi}\pbj e^{\phi})$.

The action which produces this equation of motion is a straightforward
generalization of the WZW model where the two-dimensional derivatives
$\dz$ and $\dzbar$ are replaced by $\Gp$ and $\Gtp$. The string field
theory action is
\eqn\act{\half\int\left( (e^{-\Phi} G^+ e^{\Phi})(e^{-\Phi}\Gtp e^{\Phi})
-\int_0^1 dt (e^{-t\Phi}\partial_t
e^{t\Phi})\{ e^{-t\Phi}G^+ e^{t\Phi},
e^{-t\Phi}\Gtp e^{t\Phi}\} \right) .}
Note that this action is real since $\overline{(e^{-\Phi}G^+ e^\Phi)}^R=
(\Gtp e^{\Phi})e^{-\Phi}$.
In addition to producing the correct linearized
equations of motion and three-point tree amplitude, this action contains
the non-linear
gauge invariance,
\eqn\gauge{\delta e^{\Phi}= (G^+\Lambda) e^\Phi  + e^\Phi (\Gtp\bar\Lambda) ,}
which generalizes the linearized gauge invariance
$\delta\Phi=\Gp\Lambda+\Gtp\bar\Lambda$.

\newsec{ Review of the N=2 Description of the Superstring}

As was described in reference \bv, an N=2 description of the superstring
can be obtained by ``embedding'' the critical N=1 string into
a critical N=2 string. The resulting N=4 superconformal generators
(after twisting) consist of
$$L=L_{RNS},\quad G^+=j_{BRST},\quad \Gtp =\eta,\quad G^-=b,\quad
\tilde G^-=bZ ,$$
\eqn\RNS{J^{++}=c\eta,\quad J=cb+\eta\xi,\quad J^{--}=b\xi,}
where $L_{RNS}$ is the RNS stress-energy tensor (including the N=1 ghosts),
$Q_{RNS}=\int j_{BRST}$, $Z=\{Q,\xi\}$, and the N=1 ghosts are fermionized
as $\gamma=\eta e^{\phi}$ and $\beta=(\dz\xi)
e^{-\phi}$.\vafa
Note that the ``large'' hilbert space is necessary for the N=2
description since the zero mode of $\xi$ explicitly appears in
the superconformal generators.

As in the N=2 string for (2,2) self-dual Yang-Mills, the physical
vertex operators for the superstring satisfy
$\Gtp\Gp\Phi=0$ and are subject to the gauge invariances
$\delta\Phi=G^+\Lambda+\Gtp\bar\Lambda$ (note that the N=2 vertex operator
$\Phi$ is related to the N=1 vertex operator $A$ by $\Phi=\xi A$,
so $Q A=0$ implies that $\Gtp\Gp \Phi =0$). However
a crucial new feature for the superstring is that $G^+$ and
$\Gtp$ have trivial cohomology (i.e., $G^+ \Phi=0$ implies that
$\Phi=G^+(\xi Y \Phi)$ and $\Gtp\Phi=0$ implies that
$\Phi=\Gtp(\xi\Phi)$).
As was shown in reference \vafa, this means that each physical state
of the superstring is represented by an infinite ladder of vertex
operators, $V_n$, where $n$ labels the N=1 picture.
Adjacent steps on the ladder
are related by $G^+ V_n =c_n \Gtp V_{n+1}$ for some constant $c_n$.

The main advantage of the N=2 description of the superstring is that
it can be made manifestly
SO(3,1) super-Poincar\'e invariant
for any compactification which preserves four-dimensional
spacetime-supersymmetry
(this is not possible in the
N=1 description since the N=1 fermionic generator is not
GSO projected). By finding a field redefinition from
RNS variables to GS variables, all superconformal
generators and vertex operators can be expressed in manifestly
super-Poincar\'e invariant notation. This field redefinition takes the
four-dimensional RNS variables $x^m$, $\psi^m$, $c$, $b$, $\xi$, $\eta$,
and $\phi$ into the four-dimensional GS variables $x^m$, $\t^\a$, $\tb^\ad$,
$p_\a$, $\pb_\ad$, and $\rho$ ($p_\a$ and $\pb_\ad$ are conjugates
to $\t^\a$ and $\tb^\ad$, and $\rho$ is a chiral boson similar to $\phi$ of
the N=1 ghost sector).
As in the RNS formalism, the internal six-dimensional variables of the
GS superstring will be described by a $c=9$ N=2 superconformal field theory.

Under this field redefinition, the twisted
N=2 superconformal generators of equation \RNS get mapped into the
following generators:
$$L=L_4+L_6,\quad G^+=G^+_4+G^+_6,\quad G^-=G^-_4+G^-_6,\quad J=J_4+J_6$$
where
\eqn\GSf{L_4=\half\dzm x^m \dzm x_m +
p_\a\dzm \t^\a +  \pb_\ad \dzm\tb^\ad +\half\dzm\rho\dzm\rho+\half
\partial^2\rho}
$$G^+_4=e^{\rho} (d)^2 , \quad
G^-_4=e^{-\rho} ( \bar d)^2, \quad
J_4=-\dzm\rho, $$
$d_\a=p_\a+i\tba\dzm x_{\a\ad}-\half(\tb)^2\dzm\t_\a
+{1\over 4}\t_\a \dzm (\tb)^2,$
$ \bar d_\ad= \bar p_\ad
+i\ta\dzm x_{\a\ad}-\half(\t)^2\dzm\tb_\ad
+{1\over 4}\tb_\ad \dzm (\t)^2, $
$(d)^2$ means
$\epsilon^{\a\b} d_\a d_\b$, and $[L_6,G^+_6, G^-_6, J_6]$ form a
$c=9$ N=2 superconformal field theory.

The remaining N=4 superconformal generators can be constructed
from these N=2 generators in the usual way. The only such
generator which will
be needed for the string field theory is $\Gtp=\Gtpf+\Gtps$, where
$$\Gtpf=e^{\int^z J_6} e^{-2\rho}(\bar d)^2,\quad
\Gtps=e^{-\rho}\tilde G^{++}_6,$$
and $G^{++}_6$ is the pole term in the OPE of
$G_6^-$ and $e^{\int^z J_6}.$

These generators are manifestly invariant under the four-dimensional
spacetime-supersymmetry generated by
$q_\a=\int dz(p_\a-i\tba\dzm x_{\a\ad}-{1\over 4}(\tb)^2\dzm\t_\a)$ and
$\bar q_\ad=\int dz( \bar p_\ad
-i\ta\dzm x_{\a\ad}-{1\over 4}(\t)^2\dzm\tb_\ad),$
which satisy the relation $\{q_\a,\bar q_\ad\}=-2i\int dz\dz x_{\a\ad}$.
Note that the RNS picture operator, $\int dz
(\dz\phi-\xi\eta)$, is mapped into
$\int dz (\dz\rho +\half(p_\a\t^\a-\pb_\ad
\tb_\ad))$, so $q_\a$ is in the $+\half$ picture while
$\bar q_\ad$ is in the $-\half$ picture.

In certain pictures, vertex operators for the massless fields
take a particularly simple form
when written in terms of the GS variables.
The vertex operator for the four-dimensional super-Yang-Mills multiplet
is $\Phi=v(x,\t,\tb)$ where $v$ is the real superfield for the prepotential,
while the vertex operators for the Calabi-Yau scalar multiplet are
$\Phi=(\tb)^2 e^\rho \bar\Psi^j\omega_j(x,\t)$ and
$\Phi=(\t)^2 e^{-\rho}\Psi_j \obj(x,\tb) $
where
$\Psi_j$ and $\bar\Psi^j$ are the
chiral and anti-chiral primary fields for the internal $c=9$ N=2
superconformal field theory,
and $\omega_j$ and $\obj$ are chiral and anti-chiral superfields
for the scalar multiplet.

\newsec{Open Superstring Field Theory}

\subsec{Construction of the Three String Fields}

Given the results from the previous sections, an obvious guess
for constructing a superstring field theory action is to generalize
the on-shell vertex operator $\Phi$ to an off-shell string field.
However this would be incorrect since, as was shown in the previous
section, each physical state is represented by an infinite number
of vertex operators satisfying $\Gtp\Gp\Phi=0$. Although one could
restrict all bosonic vertex operators to be in the zero picture and
all fermionic vertex operators to be in the $-\half$ picture, this
would break the manifest spacetime supersymmetry.

The way to avoid this problem is to realize that although $\Gtp\Gp\Phi=0$
has infinitely many solutions, only one such solution satisfies
the more restrictive equation
\eqn\heq{(\Gp+\Gtp)\hat\Phi=0.}
It is easily shown that
up to an overall constant, this solution is
$\hat\Phi=\sum_{n=-\infty}^{\infty} V_n$ where $V_n$ are eigenvectors
of the picture operator which satisfy $G^+ V_n =-\Gtp V_{n+1}$.
Note that $(G^+ +\Gtp)^2=0$, so the solution contains the gauge
transformations
\eqn\gauget{\delta\hat\Phi=(G^+ +\Gtp)\hat\Lambda.}

Since the picture operator does not commute with spacetime supersymmetry,
it will be more convenient to expand $\hat\Phi$
in eigenvectors of the $\int\partial\rho$ operator
as
\eqn\exp{\hat\Phi=\sum_{n=-\infty}^{\infty} \Phi_n}
where $\Phi_n=e^{n\rho} F_n$ and $F_n$ carries Calabi-Yau charge $-n$.
Note that $\hat\Phi$ satisfies the reality condition of equation \conj,
so
$(\overline{\Phi_n(\sigma)})^R= \Phi_{-n}(\pi-\sigma)$.

  From equation \heq, the $\Phi_n$'s must satisfy
\eqn\eq{\Gpf \Phi_n+\Gps \Phi_{n+1}+\Gtps \Phi_{n+2}+\Gtpf \Phi_{n+3}=0}
since ($\Gpf,\Gps,\Gtps,\Gtpf$) carry $\rho$-charge ($1,0,-1,-2)$.
Under the gauge transformations parameterized by
$\hat\L=
\sum_{n=-\infty}^{\infty} \L_n$ where $\L_n$ carries $\rho$-charge
$n$,
\eqn\g{\delta\Phi_n=\Gpf \L_{n-1} +\Gps \L_n +\Gtps\L_{n+1}+
\Gtpf\L_{n+2}.}

It will now be proven that up to gauge transformations, all
$\P_n$'s can be expressed in terms of $\P_{-1}$, $\P_0$, and $\P_1$.
The proof is inductive in $m$. Suppose that $\P_p$ is known for
$|p|\leq m$ where $m$ is positive. Then for equation \eq where $n=-m-1$ or
$n=m-2$, one can solve for $\Gpf\P_{-m-1}$ and $\Gtpf\P_{m+1}$
in terms of the $\P_p$'s. But under the gauge transformations
parameterized by
$\L_{-m-2}$ and $\L_{m+3}$, $\delta\P_{-m-1}=\Gpf\L_{-m-2}$,
$\delta\P_{m+1}=\Gtpf\L_{m+3}$, and all $\P_p$'s for
$|p|\leq m$ are left unchanged. Since $\Gpf$ and $\Gtpf$ have trivial
cohomology (note that
$\Gpf (e^{-\rho}(\t)^2)$=1 and
$\Gtpf (e^{2\rho-\int^z J_6}(\tb)^2)$
=1), this means that $\P_{-m-1}$ and $\P_{m+1}$ can be expressed in terms
of the $\P_p$'s up to a gauge transformation.

So any solution to $(G^+ +\Gtp)\hat\Phi=0$ can be expressed in terms
of the three string fields $\P_{-1},$ $\P_0$, and $\P_1$. By
analyzing equation \eq for $n=-1$, 0, and 1, it is easy to check that
the remaining three string fields must satisfy the linearized equations
of motion:
\eqn\leq{\Gtpf\Gpf\P_{-1}+\Gtpf\Gps\P_0+\Gtpf\Gtps\P_1=0,}
$$(\Gtps\Gps+\Gtpf\Gpf)\P_{0}+\Gtpf\Gps\P_1+\Gtps\Gpf\P_{-1}=0,$$
$$\Gpf\Gps\P_{-1}+\Gpf\Gtps\P_0+\Gpf\Gtpf\P_1=0.$$

These equations of motion are invariant under the six independent
linearized gauge transformations:
\eqn\lg{\delta\P_{-1}=\Gpf\L_{-2} +\Gps\L_{-1}+ \Gtps\L_{0}+\Gtpf\L_1,}
$$\delta\P_{0}=\Gpf\L_{-1} +\Gps\L_{0}+ \Gtps\L_{1}+\Gtpf\L_2,$$
$$\delta\P_{1}=\Gpf\L_{0} +\Gps\L_{1}+ \Gtps\L_{2}+\Gtpf\L_3.$$

So up to these gauge transformations, there is a one-to-one
correspondence between physical states of the superstring and
fields $\P_{-1}$, $\P_0$, $\P_1$ satisfying equation \leq.

\subsec{Construction of the Superstring Field Theory Action}

In order to construct a superstring field theory action, one needs
to find a non-linear generalization of the equations of motion and
gauge invariances from \leq and \lg. Miraculously, one can guess
this non-linear generalization by analyzing the paper of Marcus,
Sagnotti, and Siegel on ten-dimensional super-Yang-Mills written
in terms of four-dimensional superfields.\mar

In this paper, it was found useful to covariantize the four-dimensional
superspace derivatives as
\eqn\pcov{\N_\a=e^{-v} D_\a e^v,\quad \Nb_\ad=\bar D_\ad,}
and the six-dimensional spacetime derivatives as
$$\Nb^j= e^{-v}(\pjb+\obj)e^v,\quad \N_j= \partial_j -\oj,$$
where $v$ is the real superfield which describes the
four-dimensional part of the gauge field, $\oj$ and $\obj$ are the
chiral and anti-chiral superfields which describe the
six-dimensional part of the gauge field ($j=1$ to 3), and
$\N_A= D_A -i\Gamma_A$ where $\Gamma_A$ is the super-connection
(note that $\oj=
i\Gamma_j$ and $e^{-v}\obj e^v=-i\Gamma^j$).

These covariant derivatives satisfy the identities
\eqn\pid{F_{\a\beta}=F_{\ad\dot\beta}=F_\a^j=F_{\ad j}=0}
where $F_{AB}=\{\N_A, \N_B]$, and transform as
$\delta\N_A= [\N_A,\sigma]$ under the gauge-transformation
\eqn\pg{\delta e^v=\bar\sigma e^v + e^v \sigma,\quad
\delta\oj= -\partial_j \sigma+[\oj,\sigma],\quad
\delta\obj= -\pjb \bar\sigma-[\obj,\bar\sigma],}
where $\bar\sigma= (D)^2\lambda$ and $\sigma=(\bar D)^2\bar\lambda$
for some $\lambda$ and $\bar\lambda$.

In terms of these four-dimensional superfields, the non-linear
equations of motion for ten-dimensional super-Yang-Mills are:
\eqn\peq{2\{\N^\a, W_\a\}=F_j^j, \quad
2\{\N^\a, F_{\a j}\}=\epsilon_{jkl} F^{kl},\quad
2\{\Nb^\ad, F_\ad^j\}=\epsilon^{jkl} F_{kl},}
where $W_\a=[\Nb^\ad,\{\N_\a,\Nb_\ad\}]$ is the four-dimensional
chiral field strength.

Recall that the vertex operator for the super-Yang-Mills multiplet
is simply $\Phi=v$, while for the Calabi-Yau massless multiplet,
it is $\Phi=e^{\rho}(\tb)^2\bar\Psi^j\oj$
and
$\Phi=e^{-\rho}(\t)^2\Psi_j\obj$. The massless
ten-dimensional super-Yang-Mills fields are therefore described by
$\P_{-1}
=e^{-\rho}(\t)^2\psi_j\obj$,
$\P_0=v$, and $\P_1=
e^{\rho}(\tb)^2\bar\psi^j\oj$.

Since $\Gpf=e^{\rho}(d)^2$ and $\Gtpf={1\over 6}
e^{-2\rho}\epsilon^{jkl}
\psi_j
\psi_k\psi_l(\bar d)^2$ in a flat background, it is
natural to covariantize these operators
to
\eqn\scov{\Gf=e^{-V}\Gpf e^V,\quad \Gtf=\Gtpf}
where $V\equiv\Phi_0$ (so for massless excitations,
$d_\a\to e^{-v} d_\a e^v$ and $\bar d_\ad
\to \bar d_\ad$). Similarly, since
$\Gps=\psi_j\dz\bar x^j$ and
$\Gtps=\half e^{-\rho}\epsilon^{jkl}\psi_j\psi_k\dz x_l$ in a flat
background, it is natural to covariantize these operators to
$$\Gs=e^{-V}(\Gps+\bar\Omega)e^V \quad\Gts=\Gtps-\Omega$$
where $\Ob\equiv\Gpf\P_{-1}$ and $\O\equiv
\Gtpf\P_1$ (so for massless excitations,
$\dz \bar x^j\to\dz\bar x^j-\obj$ and $\dz x_j\to\dz x_j+\oj$).

Like their point-particle counterparts in \pid, these covariantized operators
satisfy the identities:
\eqn\sid{\{\Gf,\Gf\}=\{\Gtf,\Gtf\}=\{\Gf,\Gs\}=\{\Gtf,\Gts\}=0}
and transform as $\delta{\cal{G}}_A=[{\cal{G}}_A,\Sigma]$ under
the gauge transformations
\eqn\sg{\delta e^V=\bar\Sigma e^V +e^V \Sigma,\quad
\delta\O=-\Gtps\Sigma
+[\O,\Sigma],\quad
\delta\Ob=-\Gps\bar\Sigma-[\Ob,\bar\Sigma]}
where $\bar\Sigma=\Gpf\L_{-1}$ and $\Sigma=\Gtpf\L_2$.

A natural string generalization of the point-particle equations of
motion in equation \peq is
\eqn\seq{\{\Gf,\Gtf\}=-\{\Gs,\Gts\},}
$$
2\{\Gf,\Gts\}=-\{\Gs,\Gs\},\quad
2\{\Gtf,\Gs\}=-\{\Gts,\Gts\}.$$
These equations can be combined with the identities of equation \sid
to imply that
\eqn\f{(\Gf+\Gs+\Gtf+\Gts)^2=0,}
which is the natural generalization of $(Q+A)^2=0$ for the
Chern-Simons-like action.

In addition to the gauge invariances of equation \sg, the equations
of motion implied by \f are also invariant under the following gauge
transformations:
\eqn\add{\delta e^V=e^V (\Gs\L_0 +\Gts\L_1),}
$$ \delta\O=\Gtpf(\Gf\L_0 +\Gs\L_1),\quad
\delta\Ob=\Gpf(e^V( \Gts\L_0 +\Gtf\L_1)e^{-V}).$$
Unlike the gauge transformations of equation \sg,
these gauge transformations have no super-Yang-Mills counterpart since
there is no massless contribution to $\L_0$ or $\L_1$.

Finally, a superstring field theory action which yields these equations
of motion can be constructed by comparing with the following point-particle
action of reference \mar for ten-dimensional super-Yang-Mills:
\eqn\pac{\half\int d^{10}x [\,\, -2\int d^2 \t \, W^\a W_\a  }
$$+
\int d^4 \t \left( (e^{-v}\pjb e^v)(e^{-v}\pj e^v)-\int_0^1 dt
(e^{-tv}\partial_t e^{tv})
\{ e^{-tv}\pjb e^{tv},
e^{-tv}\pj e^{tv}\} )\right)
$$
$$+2\int d^4\t\left( (\pj e^{-v})\obj e^v+
 e^{v}\oj(\pjb e^{-v})+ e^{-v}\obj e^v\oj\right)$$
$$ +\int d^2\t \epsilon^{jkl}
(\oj\partial_k\omega_l+{2\over 3}\oj\omega_k\omega_l)+
\int d^2\tb \epsilon_{jkl}
(\obj\bar\partial^k\obl-{2\over 3}\obj\obk\obl)\,\,].$$

The superstring generalization of this action is:

\eqn\sa
{\half\int [\,\, (e^{-V}\Gpf e^V)(e^{-V}\Gtpf e^V)-\int_0^1 dt
(e^{-tV}\partial_t e^{tV})
\{ e^{-tV}\Gpf e^{tV},
e^{-tV}\Gtpf e^{tV} \} }
$$+(e^{-V}\Gps e^V)(e^{-V}\Gtps e^V)-\int_0^1 dt
(e^{-tV}\partial_t e^{tV})
\{ e^{-tV}\Gps e^{tV},
e^{-tV}\Gtps e^{tV}\} $$
$$+
2\left( (\Gtps e^{-V})\Ob e^V+
 e^{V}\O(\Gps e^{-V})+ e^{-V}\Ob e^V\O\right)$$
$$-(\O\Gtps\P_{1}-{2\over 3}\O\O\P_{1})+
(\Ob\Gps\P_{-1}+{2\over 3}\Ob\Ob\P_{-1})\,\,].$$

The only subtle part of this generalization is that unlike the
point-particle action, a WZW term is used for both the four-dimensional
and six-dimensional parts of the string action for $V$. Note also that
chiral $F$-terms in the point-particle action are annihilated by
$\Gtpf$ in the string action and, because $\Gtpf$ has trivial
cohomology, can be turned into $D$-terms by pulling $\Gtpf$ off one
of the $\O$'s. Since $\Gtpf$ is the inverse of the $\xi$ zero mode,
turning $F$-terms into $D$-terms is like going from the small
to the large RNS hilbert space.

To show that this superstring field theory action is correct, one should
check that its linearized equations of motion and gauge invariances
reproduce the on-shell conditions of physical vertex operators, and that
the cubic term in the action produces the three-point tree-level
scattering amplitude.

It is straightforward to show that the linearized part of
\seq, \sg, and \add, reproduce \leq and \lg (the
$\L_{-2}$ and $\L_{3}$ gauge transformations act trivially
on the string fields), and therefore define
the on-shell conditions for the physical vertex operator $\hat\Phi$
which is constructed out of $\P_{-1}$, $\P_0$, and $\P_1$.

The cubic term in the superstring field theory action of \sa is:
\eqn\cubicterm{\half\int [\,{-1\over 3}V ( \{\Gpf V, \Gtpf V\}+
 \{ \Gps V, \Gtps V\})}
$$- V(\{\Gtps V,\Ob\}
+\{\Gps V,\Ob\}
+ 2\{\O, \Ob\})
+{2\over 3}(\O\O\P_{1}-\Ob\Ob\P_{-1})\,\,].$$
Suppose that the
three-point amplitude involves vertex operators, $\Phi(z_r)$,
which contain
no Calabi-Yau charge in the zero picture. Then
$V(z_r)=\Phi(z_r)$ and $\O(z_r)=\Ob(z_r)=0$. It is easy to check that in
this case, \cubicterm reproduces the
three-point amplitude of \three. But since the coefficients
in \cubicterm are restricted by the requirement of on-shell
gauge invariance under
the transformations of \lg, this is
enough to prove that \cubicterm
gives the correct three-point amplitude even
when $\O$ and $\Ob$ are non-zero.

\newsec{Conclusion}

In this paper, a WZW-like field theory action was constructed
for open strings with critical N=2 superconformal invariance.
For the N=2 string which describes (2,2) self-dual Yang-Mills,
this field theory action generalizes the point-particle action for
the scalar field of Yang. For the N=2 string which describes the
superstring in a Calabi-Yau background, the action generalizes the
point-particle action for ten-dimensional super-Yang-Mills written in terms
of four-dimensional superfields.

In both of these string field theory actions, only the U(1)-neutral
string fields were considered, which correspond to the matter sector
of the field theory. However it should also be possible to consider
U(1)-charged string fields which represent the ghost sector
of the field theory. Studying the ghost contributions to these actions
would be interesting since for (2,2) self-dual Yang-Mills, there
is a conjecture that topological supersymmetry relates the
matter and ghost sectors.\vafa Furthermore,
there is a conjecture
that adding four bosonic and four fermionic degrees of freedom to the
light-cone is useful for constructing covariant actions of
supersymmetric field theories.\ref\fo{W. Siegel,
Int. J. Mod. Phys. A4 (1989) 1827.}
Although the superstring field
theory action in this
paper involves adding three bosons ($x^+$, $x^-$, and $\rho$)
and six fermions ($\tb^\ad$, $\pb_\ad$, $\t^2$, and $p_2$)
to the light-cone, two of the fermions could
be bosonized to produce a fourth boson.

A second important result of this paper was the discovery that
all physical states of the superstring are uniquely represented
by three string fields, $V$, $\Omega$, and $\Ob$, which generalize
the real, chiral, and anti-chiral superfields of four-dimensional
superspace.
In terms of $V$, $\O$, and $\Ob$, it was straightforward to generalize
from the ten-dimensional super-Yang-Mills action to an open
superstring field theory action. Maybe these three string fields can
be used
to generalize other properties of four-dimensional supersymmetric
field theories (perhaps even non-perturbative properties)
from point-particle
language to the superstring.

{\bf Acknowledgements:} I would like to thank Warren Siegel and Cumrun
Vafa for many useful conversations, and Rutgers University for their
hospitality. This work was financially supported by the
Conselho Nacional de Pesquisa.

\listrefs
\end